\documentstyle[11pt,newpasp,twoside,epsf]{article}
\def\edcomment#1{\iffalse\marginpar{\raggedright\sl#1\/}\else\relax\fi}
\reversemarginpar

\begin{document}

\newcommand{\HI}{\mbox{H\,{\sc i}}}
\newcommand{\am}[2]{$#1'\,\hspace{-1.7mm}.\hspace{.0mm}#2$}

\title{ce-61: a Tidal Dwarf Galaxy in the Hercules cluster?}
 \author{W. van Driel$^1$}
\affil{1. Observatoire de Paris, GEPI, Meudon, France }
\author{P.-A. Duc$^2$, P. Amram$^3$, F. Bournaud$^2$,
C. Balkowski$^1$, V. Cayatte$^1$, 
J. Dickey$^4$, H. Hern\'andez$^5$, J. Iglesias-P\'aramo$^6$, K. O'Neil$^6$, P. Papaderos$^7$, J.M. V{\'{\i}}lchez$^8$ }
\affil{2. CEA, Service d'Astrophysique, Gif-sur-Yvette, France;\,\,\,\,\,
3. Laboratoire d'Astrophysique de Marseille, Marseille, France,
4. Dept. of Astronomy, Univ. of Minnesota, Minneapolis, MN, USA,
5. Arecibo Observatory, Arecibo, PR, USA,
6. NRAO, Green Bank, WV, USA,
7. Universitäts-Sternwarte, G\"ottingen, Germany,
8. Instituto de Astrofísica de Andalucía (CSIC), Granada, Spain }

\begin{abstract}
A candidate Tidal Dwarf Galaxy, ce-61, was identified in the merger system IC 1182
in the Hercules supercluster. The multi-wavelength data we obtained so far do not
prove, however, that it is kinematically detached from the IC 1182 system and gravitationally 
bound.        
\end{abstract}

\vspace*{1cm}
\noindent
The Hercules supercluster (D=150 Mpc, H$_0$=75) is one of the most massive structures in the nearby 
Universe. We studied (Iglesias-P\'aramo et al. 2003) 22 \HI-selected galaxies in this cluster, 
from the blind VLA \HI\ survey by Dickey (1997), obtaining:
deep CCD $B$, $V$ and $I$-band  surface photometry of 10 galaxies, 
optical spectroscopy of 8 of these,
Arecibo \HI\ observations of all 22 galaxies and
H$\alpha$ line Fabry-Perot observations of the IC 1182 merger system.

Based on these multi-wavelength observations, the object ce-61 was identified as a candidate 
Tidal Dwarf Galaxy (TDG) in a tidal tail of the peculiar IC 1182 system.
IC 1182 ($B_T$=15.4, $V$=10,223 km/s) shows several characteristics 
typical of a merger system, e.g., a blue optical jet-like structure 
towards the East and tidal debris towards the NW, and an 
extended \HI\ distribution with two tidal tails, towards the E and the NW. 
The candidate TDG  ce-61 ($M_B= -18.24$  mag) lies at the tip of the eastern optical/\HI\
tail, at about \am{1}{5} (65 kpc) projected distance from the centre of the parent system. 
Its CCD image shows two distinct peaks and the maximum in the \HI\ tail coincides with the 
easternmost optical peak. Its metallicity (8.41) is on the high side for a dwarf galaxy of its luminosity, 
but typical for a TDG. 
It is a very gas-rich system, with an estimated  $M_{HI}$/$L_{B}$ ratio of 6 $M_{\odot}$/L$_{\odot,B}$; 
its \HI\ line width is about 220 km/s.
CO line observations (Braine et al. 2001) show about 7 10$^9$ $M_\odot$ of H$_2$ in a resolved 
distribution in IC 1182,  but none was detected in ce-61, putting an upper limit to its H$_2$ mass of about 
6 10$^7$ $M_\odot$.

In order to study whether the TDG candidate is already kinematically 
detached from the IC 1182 system and gravitationally bound, we 
obtained and analyzed H$\alpha$ line Fabry-Perot observations and found
(Bournaud, Duc \& Amram 2003, in prep.) that:
(1) the brightest knot at the tip of the tail, coinciding with ce-61, seems to be 
kinematically detached from the overall velocity field along the tail (which is 
governed by streaming motions); the offset is  about 30 km/s. However, 
our numerical simulations show that this offset can be consistent with a projection effect 
along the line of sight (with the tail, seen edge-on, being bent in 3D space), 
(2) along direction 2 (see Fig.), there is a hint of an internal velocity gradient, of  70 km/s maximum, 
associated with one of the knots. We lack the spatial resolution to confirm it, however.

% \vspace*{-0.0cm}
\begin{figure}[h]
 \plotone{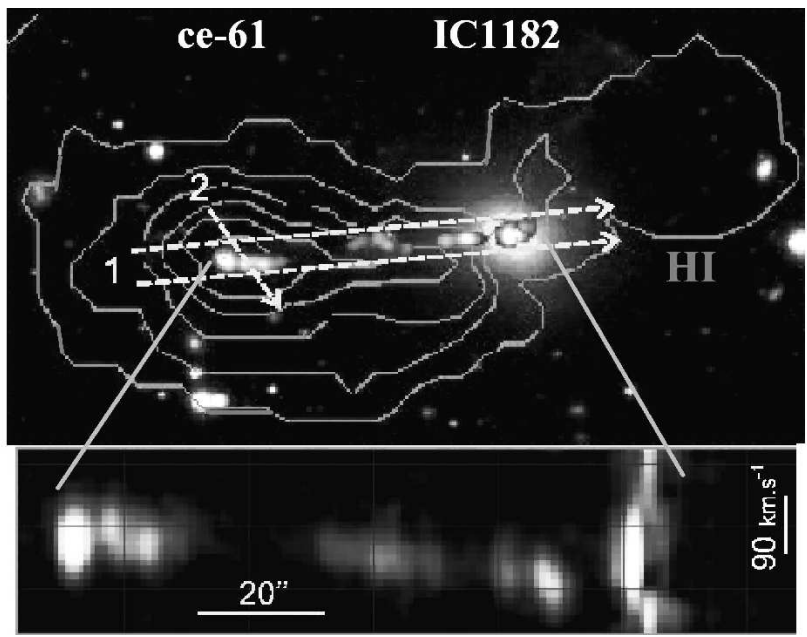}
      \caption{H$\alpha$ line Fabry-Perot observations of the candidate TDG ce-61 in the merger system 
IC 1182: (top)  the clumpy distribution of the H$\alpha$ gas in the tidal tail of IC 1182, superimposed 
on a $V$-band  optical image of the system + \HI\ column density contours; (bottom) an  H$\alpha$ line 
position-velocity diagram along the tidal tail (following the lines marked `1' in the top panel).
  }
\end{figure}
\vspace*{-0.1cm}

% \acknowledgments{I thank everyone ....}

\end{document}